\documentclass[9.5pt,journal,compsoc]{IEEEtran}

\usepackage[most]{tcolorbox}
\usepackage[utf8]{inputenc}
\usepackage{array,multirow,etoolbox,graphicx,subcaption,fancyhdr,tabularx,longtable}
\usepackage{float}
\usepackage{graphicx}
\usepackage{amsmath} 
\usepackage{booktabs}
\usepackage{mathtools}
\usepackage{listings}
\usepackage{xcolor}
\usepackage{hyperref}
\newtcolorbox{TcolorBox}[1]{fonttitle=\bfseries,title=#1}
\usepackage{csquotes}
\usepackage[euler]{textgreek}
\usepackage{microtype}
\usepackage{tablefootnote}
\usepackage{threeparttable}
\usepackage{fvextra}
\usepackage{adjustbox} 

\newlength{\imagewidth}
\usepackage{amsmath,amsfonts}
\usepackage{algorithmic}
\usepackage{graphicx}
\usepackage{textcomp}
\usepackage{xcolor}
\usepackage{url}
\usepackage{microtype}

\usepackage{xurl} 
\usepackage{tabularx}
\usepackage{caption}
\usepackage{subcaption}
\usepackage{xspace}

\usepackage{enumitem}
\newlist{steps}{enumerate}{1}
\setlist[steps, 1]{label = Step \arabic*:}

\usepackage{xspace}
\usepackage{dirtytalk}
\usepackage{xcolor}

\newcommand{\company}{Meta\xspace}
\newcommand{\Meta}{Meta\xspace}
\newcommand{\testfailurebot}{TFMB\xspace}
\newcommand{\benchmark}{TF Benchmark\xspace}
\newcommand{\patchgen}{PatchGen\xspace}
\newcommand{\agent}{Engineering Agent\xspace}

\newcommand{\diff}{diff\xspace}
\newcommand{\diffs}{diffs\xspace}

\newcommand{\diagnostics}{static analysis tools\xspace}
\newcommand{\Diagnostics}{Static Analysis Tools\xspace}

\usepackage{xspace}
\usepackage{dirtytalk}
\usepackage{xcolor}

\definecolor{teal}{RGB}{0,128,128}

\newcommand{\DefMacro}[2]{\expandafter\newcommand\csname rmk-#1\endcsname{#2}}
\newcommand{\UseMacro}[1]{\csname rmk-#1\endcsname}
\newcommand{\rom}[1]{\uppercase\expandafter{\romannumeral #1\relax}}

\usepackage{xcolor}
\usepackage{xspace}

\newcommand{\etal}{\hbox{\emph{et al.}}\xspace}

\newcommand{\ie}{\hbox{\emph{i.e.}}\xspace}

\PackageWarning{TODO:}{X}
\PackageWarning{TODO:}{X\%}

\usepackage{tcolorbox}

\begin{document}
\title{Agentic Program Repair from Test Failures at Scale: A Neuro-symbolic approach with static analysis and test execution feedback}

\author{Chandra Maddila, Adam Tait, Claire Chang, Daniel Cheng, Nauman Ahmad, Vijayaraghavan Murali, Marshall Roch, Arnaud Avondet, Aaron Meltzer, Victor Montalvao, Michael Hopko*, Chris Waterson, Parth Thakkar, Renuka Fernandez, Kristian Kristensen, Sivan Barzily, Sherry Chen, Rui Abreu, Nachiappan Nagappan, Payam Shodjai, Killian Murphy, James Everingham, Aparna Ramani, Peter~C. Rigby 
\IEEEcompsocitemizethanks{\IEEEcompsocthanksitem All the authors are with Meta Platform Inc, Menlo Park, California, USA. *Michael Hopko was with Meta when the work was done. P. C. Rigby is also with the Department of Computer Science and Software Engineering, Concordia University, Montr{\'e}al, Qu{\'e}bec, Canada.\protect\\
Corresponding Author E-mail: cmaddila@meta.com}
}

\markboth{Transactions on Software Engineering,~Vol.~X, No.~Y, July~2025}%
{Shell \MakeLowercase{\textit{Maddila \etal}}: Agentic Program Repair from Test Failures}

\IEEEtitleabstractindextext{

\textbf{Aim.} With the advent of LLMs, sophisticated agentic program repair has become viable at large organizations with large codebases. In this work, we develop an \agent that fixes the source code based on test failures at scale across diverse software offerings internally.

\textbf{Method.} Using Llama as the base, we employ the ReAct harness to develop an agent. We start with a test failure that was triaged by a rule-based test failure bot. We then setup an agentic harness and allow the agent to reason and run a set of 15 actions from reading a file to generating a patch. We provide feedback to the agent through static analysis and test failures so it can refine its solution. Once the validations are passing, we use a separate LLM-as-a-Judge to ensure that the patch conforms to the standards for patch fixes at Meta. Finally, a human reviewer is notified and reviews the patch, and, if it is accepted, lands the change in our mono repo.
This work involved large research and production engineering teams. 

\textbf{Benchmark Findings.} We curated offline benchmarks for our patch generator, the Engineering Agent loop, and the LLM-as-a-Judge. In offline evaluations we found that a specialized 70B model, that is internally fine-tuned for patch generation using the search-replace format, is highly competitive with the much larger but vanilla Llama-405B.  We also found that the diff format impacts performance with the "search-and-replace" format outperforming the standard unified diff format. In an ablation study, we found that the ReAct harness (neural model) benefited from the symbolic information from static analysis tools and test execution traces. Our ablations allowed us to strike a balance between the solve rate and error rate vs the cost and latency. The balanced model has a benchmark solve rate of 42.3\% using an average 11.8 feedback iterations. We then experimented with the Engineering Agent in production.

\textbf{Production Findings.}  In a three month period, 80\% of the generated fixes were reviewed, of which 31.5\% were landed (25.5\% of the total number of generated fixes).

\textbf{Feedback from Engineers.} We used open coding to extract qualitative themes from engineers' feedback. We saw positive feedback in the form of quick approvals, gratitude, and surprise. From the negative feedback we modified our production system. For example, feedback regarding test flakiness forced us to develop a more isolated environment for the agent to run in. We also found mixed feedback when the \agent's solution was partially correct and it served as a good starting point for the human reviewer to create a final solution.

}

\IEEEdisplaynontitleabstractindextext

\IEEEpeerreviewmaketitle

\maketitle

\section{Introduction}

Repairing programs without manual intervention has been one of the ultimate goals in software engineering. Automated Program Repair (APR) has been long been studied in the software engineering and programming language community \cite{Koyuncu_2020, huang2023surveyautomatedprogramrepair, 10.1145/3318162}. This has spanned the spectrum from formal verification to the more recent advances in using language models \cite{zhang2024systematicliteraturereviewlarge}. 
Traditionally, APR systems were designed to operate within a narrow scope, relying on rule-based systems and test suites to identify and fix errors \cite{7816488, 10.1007/978-3-030-11245-5_4, Frenkel2022}. However, with the advent of machine learning methods, APR has transcended its conventional boundaries.

Automated program repair has recently undergone a profound transformation with the integration of statistical and deep learning-based models drastically enhancing the capabilities of APR systems. This has enabled them to perform complex tasks such as fault localization, patch generation, and ranking \cite{zhang2024systematicliteraturereviewlarge, rondon2025evaluatingagentbasedprogramrepair, cheng2025agenticbugreproductioneffective, 222607}. Moreover, recent breakthroughs in agent-based systems have demonstrated the potential of large language models (LLMs) in performing end-to-end software engineering tasks in intricate environments. With the advent of LLMs we find their use in program repair to be a viable opportunity to deploying them at scale in industry.

\begin{figure*}
    \includegraphics[width=2\columnwidth]{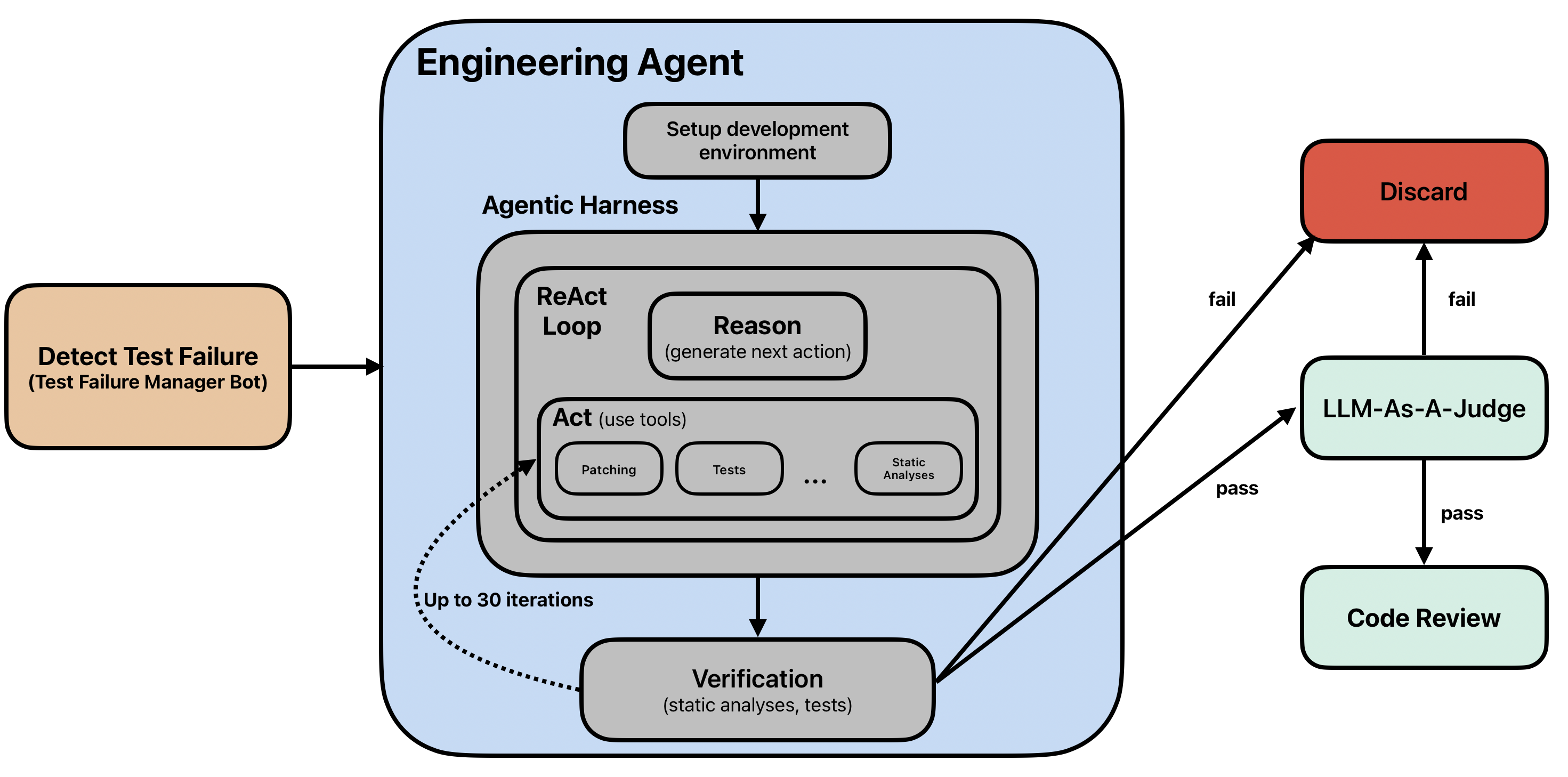}
    \caption{\agent Flow. We begin with a test failure that is triaged by the Test Failure Management Bot. The \agent then sets up the development environment using a ReAct harness performs a series of actions from reading files to generating a patch. Each patch is verified by static analysis and tests. After it passes verification it is sent to the LLM-as-a-Judge to determine if the patch is similar to acceptable fixes. If it passes this judge it is sent to a human engineer for code review. The human engineer is provided with the \agent trajectory and evidence that the verification has passed, so they can understand the ``reasons" behind the patch. The engineer can accept, modify, or reject the fix.}
    \label{fig:pipeline}
\end{figure*}

In this paper, we start from a failing test and create an agent that generates a code patch that passes the test. Since there are many possible techniques to do this we provide an overview of our \agent in Figure~\ref{fig:pipeline}. The starting point is a failing test. This test is picked up by a rule-based test failure management bot \testfailurebot and after performing some checks the failure is passed to our \agent. The \agent first sets up the development environment to replicate the failure. An agentic harness then provides an prompt, and tasks specific instruction including stack traces to the \agent. A ReAct harness ~\cite{yao2023react}  ``reasons'' to generate actions, such as running the test, searching for a file, and creating a patch. After a sequence of actions, a verification step is run which includes static analysis, testing, and an LLM based judge. The verification results are fed back into the agentic loop, and new actions can be taken. Ultimately the code change that results in a passing test is sent for code review to the author of the diff that broke the test or the test author.

Our \agent required substantial research and a large engineering team to put this research into production at scale. We summarize our contributions in terms of both industrial practice and research into agentic program repair: 

\begin{enumerate}

\item We develop an agentic harness including system prompt, task-specific instructions, and actions. {\bf Result.} After setting up the environment, we used the ReAct harness to perform "reasoning" and have a possible set of 15 actions that the \agent can perform. Unlike prior work that focused on test order dependencies~\cite{GoogleAgentProgramRepair}, a test can fail for many reasons resulting in a greater set of actions. An important aspect is that the test and static analysis act as verification with feedback for the next ReAct harness. This ensures that the \agent does not create a solution that introduces other problems. We develop an internal benchmark to evaluate \agent offline.

\item Using ablation, we compare the offline benchmark results to understand how important symbolic information is for a Neural model. {\bf Result.} The ReAct agent on its own has 28.5\% compared to ReAct with static and test execution information to 42.3\%. The stochastic nature of LLM make multiple trials necessary, and improves our best model to 61.0\%. This has important implications showing that neuro-symbolic information outperforms pure neural models. 

\item We experiment with the patch formats to find the one most amenable to LLM agents. {\bf Result.} The standard unified diff format confuses agents LLMs, and it will arbitrarily generate line numbers that may not even exist. Instead, we use the ``search-and-replace" diff format which is a more natural text format for LLMs. We validate that the diff format dramatically impacts the accuracy of the model. This has important research implications for AI code generation work. 

\item Before deploying \agent in a production environment, we trained an LLM-as-a-Judge to determine the quality of the generated patch to ensure that we did not provide low quality fixes to engineers that would reduce trust in our production system. {\bf Result.} Engineers manually labeled \agent fixes as appropriate or inappropriate. Using this benchmark, we calibrated the judge to have high precision for low quality samples, .867 precision, because the goal is to eliminate diffs that pass the test in an inappropriate manner. We want to avoid wasting engineering's effort or worse, having the code landed in the repo that is subpar.

\item We designed and deployed \agent for use internally at scale. {\bf Result.} Over a 3 month trial period, 80\% of the AI-generated fixes received a human review and 25.5\% were landed. 

\item We perform a qualitative open coding analysis of feedback from engineers using \agent. {\bf Result.} Our findings led to refinements in our deployed system and also emphasized where engineers feel additional research is needed. The major positive themes are that it saves engineer effort, it provides a starting point for a fix, and it initiates a discussion of the underlying code problems that lead to removing the technical debt that confused the \agent. The main negative points were that it is often difficult to find expert reviewers for agentic diffs, that tool actions were missing, and that solutions were only partially correct. 

\end{enumerate}

This paper is organized as follows. In Section~\ref{sec:agentOverview} we provide a comprehensive overview of our agent, which is illustrated in Figure~\ref{fig:pipeline}. In Section~\ref{secTestFailureBot}, we introduce the rule-based Test Failure Management Bot that collects information about test failures and runs bisection and passes this information to the agent. In Section~\ref{harness}, we describe our agentic harness including the actions and prompts that are provided to the agent. We also discuss our offline benchmark for the agent. In Section~\ref{subsec:patching}, we discuss the sub-stage of generating a patch and the separate benchmark we curated to evaluate patch generation. In Section~\ref{sec:ValidationMethod}, we discuss the static analysis and test execution validators that we run. We also introduce our LLM-as-a-Judge which ensures that patches are acceptable to \Meta's engineers. In Section~\ref{evaluation-method}, we describe the experiment's context, the offline evaluation methods, and the rollout and evaluation method for the production system. In Section~\ref{secOfflineResults}, we present our offline results. In Section~\ref{secProduction}, we describe the design of the production system, usage of the system, and discuss the feedback engineers provided. In Section~\ref{sec:threats}, we discuss threats to validity. In Section~\ref{sec:literatureAndDiscussion}, we position our work in the research literature. In Section~\ref{sec:conclusion}, we conclude the paper.

\section{\agent Overview} 
\label{method-overview}
\label{sec:agentOverview}

Figure~\ref{fig:pipeline} provides a graphical overview of our \agent.
When a test failure is detected, the rule-based Test Failure Manager Bot(\testfailurebot), the \agent acts on it to find a solution to fix the test failures. If the patches generated by the \agent pass verification then a \diff is published and sent for code review.

Essentially, the input passed to the \agent, the code produced by the \agent, and the validation mechanism can be represented as a triplet $\langle Specification, Patch, Oracle \rangle$. 

\textbf{Specification} is the input task description (text) that is passed to the \agent. Optionally, it should list the file names and line numbers (for the assisted setting). In case of \testfailurebot we include stack trace and any other signals generated by the static analysis tools when detecting test failures.

\textbf{Patch} is the patch fix generated by a human that fixes the test failures (turns the test from fail to pass). This is used as a ground truth or golden patch when comparing against the \agent-generated patches.

\textbf{Oracle} is the validation mechanism that we employ to test whether the \agent was able to solve a \testfailurebot task. Typically, it is the set of tests that were broken and flagged by the \testfailurebot as ``broken''. It is important for the LLMs to be guided by an Oracle (broken unit tests) because \textit{LLMs do not know what they do not know}. So there is a good chance that the patch generated by an LLM might not fit into the program and might not even solve the actual test failure. We call this method \textit{Oracle-Guided Patch Generation (OGPG)}.

Ultimately, the goal for the \agent is to consume the \textit{specification (input)}, produce a \textit{patch (output)}, and provide guarantees that the resulting program (original program plus patch) is passing the \textit{oracle}. 

To that end, we developed an agentic system that follows the ReAct~\cite{yao2023react} pattern. The \agent system is expected to take the \textit{specification} as input (test failure information, stack traces, blame change, etc. from the \testfailurebot) and produce a pair of $\langle thought\textsubscript{t}, action\textsubscript{t} \rangle$ at step \textit{t}. Then, the \textit{action\textsubscript{t}} will be executed in an environment that reflects a typical coding environment of an internal software engineer. Using neuro-symbolic AI techniques, the environment produces an \textit{observation\textsubscript{t}} at step \textit{t}. The \agent will consume to produce a pair of $\langle thought\textsubscript{t+1}, action\textsubscript{t+1} \rangle$ at step \textit{t+1}. An agentic harness helps with bringing all these components together and executes the \textit{agentic loop} in an environment. 

\section{Test Failure Manager Bot (\testfailurebot)}
\label{secTestFailureBot}
The input into our \agent is the Test Failure Manager Bot (\testfailurebot) which is a rule-based bot that is used internally to track and manage test failures. When a test fails in landed code, the \testfailurebot automatically files a task and assigns it to the responsible engineer or the author of the code. \testfailurebot assists in maintaining production stability because failing tests can be a symptom of larger problems in production systems.

\testfailurebot collects details about the failing test including test definition and details of the failure like a stack trace and error message. \testfailurebot attempts to bisect the code changes and identify the change that caused the test failure. The results of the code changes bisect are used to pinpoint the exact change that led to the failure and assign an owner to the failing test. Using the code changes bisect, \testfailurebot can gather multiple test failures based on root cause to reduce noise.


We reduce noise from failing tests by attempting to identify intermittent or inconsistently failing tests. Identifying low quality signals is done through several mechanisms such as sampling test results and statistical methods over historical results \cite{ProbabalisticFlakiness}.

Once \testfailurebot has collected all the information about the failing test and identified an owner, it creates a task which is backed by a database. \testfailurebot helps maintain the quality and reliability of the code base by ensuring that test failures are promptly addressed and resolved. In the next section, we discuss how \testfailurebot input is provided to the \agent.

\section{Orchestration and Agent Harness} 
\label{harness}

\begin{table*}
\centering
\resizebox{2\columnwidth}{!}{%
\begin{tabular}{l|p{9cm}|l}
\hline
\textbf{Tool} & \textbf{Description} & \textbf{Usage} \\
\hline
exit & Allows the agent to terminate execution. It takes a ``summary'' of all actions performed by the agent & \texttt{exit summary=$<$str$>$} \\
get\_diff\_details & Allows the agent to gather additional context about a diff (\diff). Each \diff contains changes to Meta's codebase, along with key metadata like a title, summary and files changed & \texttt{get\_diff\_details diff\_number=$<$str$>$} \\
get\_task\_details & Allows the agent to gather context from a task number. & \texttt{get\_task\_details task\_num=$<$str$>$} \\
read\_directory & Allows the agent to read a directory's contents  & \texttt{read\_directory directory\_path=$<$str$>$} \\
read\_file & Allows the agent to read file contents & \texttt{read\_file file\_path=$<$str$>$} \\
find\_file & Searches for a given file in the source code repository (from the source code checkout on the file system) & \texttt{find\_file file\_name=$<$str$>$} \\
go\_to\_line & Allows the agent to navigate to a particular line number in a ``file\_path''. The response contains 50 lines of code around the target ``line''. & \texttt{go\_to\_line file\_path=$<$str$>$ line=$<$str$>$} \\
search\_code & Allows for searching code against a Full Text Search index & \texttt{search\_code code\_str=$<$str$>$} \\
search\_code\_in\_file & Allows for searching code in a particular ``file\_path'' & \texttt{search\_code\_in\_file code\_str=$<$str$>$ file\_path=$<$str$>$} \\
search\_class & Allows for searching a particular class in the source code repository & \texttt{search\_class class\_name=$<$str$>$} \\
search\_method & Allows for searching a particular method in the source code repository & \texttt{search\_method method\_name=$<$str$>$} \\
search\_method\_in\_file & Allows for searching a particular method within a ``file\_path'' & \texttt{search\_method\_in\_file method\_name=$<$str$>$ file\_path=$<$str$>$} \\
search\_method\_in\_class & Allows for searching a particular method inside a class & \texttt{search\_method\_in\_class method\_name=$<$str$>$ class\_name=$<$str$>$} \\
run\_tests & Allows the agent to execute relevant tests (as specified by the \testfailurebot's test selection algorithm). This action parses the test execution output and returns the relevant text from the test execution feedback & \texttt{run\_tests} \\
edit & Takes a ``file\_path'' and ``instructions'' in natural language to generate a code change & edit \texttt{file\_path=$<$str$>$ instructions=$<$str$>$} \\
\hline
\end{tabular}
}
\caption{List of available Actions to the \agent. These range in complexity from a simple {\tt read\_directory} to a full {\tt edit} of the file to fix the failing test.}
\label{tab:Actions}
\end{table*}


The first part of the agentic harness is the system prompt (the set of actions are described in Table~\ref{tab:Actions}):

\begin{tcolorbox}[title={\textbf{Agent Harness: System Prompt}}] \label{sys_prompt}
You are an autonomous coding agent, here to provide solutions for coding issues.

You have been designed to assist with a wide range of programming tasks, from code editing and debugging to testing.

You have access to a variety of tools and commands that you can use to help you solve problems efficiently.

INSTRUCTIONS:

You are going to solve an issue on your own. You can use any command listed below to complete your task.

Remember, YOU CAN ONLY ENTER ONE COMMAND AT A TIME. You should always wait for feedback after every command.

When you're satisfied with all of the changes you've made, you can indicate that you are done by running the "exit" command.

Try different commands: If you run a command and it doesn't work, try running a different command. A command that did not work once will not work the second time unless you modify it.
\end{tcolorbox}

The specification of the task is passed from \testfailurebot and has the following form:

\begin{tcolorbox}[title={\textbf{Agent Harness: Task-specific Instructions}}] \label{task_specific_instruction}
You are currently trying to complete this task:

Run Output from Test Failure:

Stack trace 1:

There was 1 failure:

1) [...test name...][...test tool failure details...]

Backtrace:

  [...backtrace output...]

\end{tcolorbox}

With these constraints, the agent loops through a series of $\langle thought, action \rangle$ pairs. The agentic harness relies on a Large Language Model (LLM) to generate a new thought and action at each iteration of the ReAct loop. At every step, the agent decides on what the next action should be based on the thoughts it generated at the current step. Each action consists of a Unix-like tool name and its associated arguments. 

The \agent has access to 15 different tools shown in Table~\ref{tab:Actions}. The set of tools offer interfaces to existing internal developer tools. Most of these tools are designed to work on our large code base and in coordination with other internal tools. For example, the \agent has many code search tools because employees also have access to code search to navigate our large code base. The tool actions accept zero or more keyword arguments as shown in the following example:

\begin{tcolorbox}[title={\textbf{Agent Harness: Example Action}}]
\texttt{read\_file file\_path='xyz.py'}
\end{tcolorbox}

Upon deciding on the next action, the agent harness then parses the tool name, executes it against the development environment. The output of each tool execution is called an \textit{observation}. While working on a task, the agent maintains its {\it trajectory} which is a history of all prior thoughts, actions, and observations to guide the LLM with future decision making in its persistent memory. Therefore, all the observations and tool executions from the previous steps can influence the agent's thought process and subsequent actions at the next step.

Finally, when the agent is done working on a task, it calls the \texttt{exit} action to terminate. We also enforce a combination of a timeout and max iteration limit to ensure the agent always exits gracefully within the budget constraints enforced for its execution.

\subsection{Agent Benchmark} \label{sec:agentBenchmark}

Benchmarking and evaluation are essential to guiding the improvements in a wide input and solution space. Benchmarks are especially vital for understanding the performance of the current agentic system, comparing different models, and identifying regressions. Software engineering is a creative undertaking with an unlimited input and solution space. We attempt to capture the enormity of the distribution by building benchmarks of human authored examples. Each example is built from the triplet $\langle Specification, Patch, Oracle \rangle$. 

To evaluate the \agent on \testfailurebot tasks, we constructed \benchmark. To source the examples, we mine the database of tasks created by the \testfailurebot. We employ the following heuristics to curate the \benchmark:
\begin{itemize}
    \item \testfailurebot tasks created in the last 90 days
    \item A human has solved the test failure and landed changes
    \item Human solution was part of the web, app and systems code
    \item Human solution changed source but not test files, build, configuration, and generated-only changes
\end{itemize}


After applying these filters to \testfailurebot tasks, we apply a series of automated benchmark validations. First, all the oracle tests are evaluated without the golden human-authored code changes. The test results are verified to ensure they continue to fail. Examples with passing tests are discarded. The environment without the golden human-authored code changes becomes the environment on which the \agent is evaluated. Second, all the oracle tests are evaluated with the golden human-authored code fix. If the test results are passing then we conclude that the oracle tests are accurate and reliable. These automated validations are re-affirmed before each evaluation run.

We annotate each example in the \testfailurebot benchmark to ensure test quality. We review the test definition ensuring code coverage of the golden human-authored code changes.
The final benchmark consists of 123 real test failures that occurred in two large mono repositories across 15 programming languages. Our benchmark is of a similar size to other industrial projects~\cite{GoogleAgentProgramRepair}.

\section{Patch Generation}
\label{subsec:patching}

Patch generation is a critical step of \agent, as the ultimate task of the agent is to edit code and generate a fix. The patch generating model, or ``patcher'', is designed as a sub-agent with a different goal than the orchestrator. It is not exposed to the set of unix tools, rather it is prompted with the task of generating a patch, given contextual code and an instruction. It is shown a summarized history of the orchestrator's trajectory in order to understand the overview of the task.

The format in which the patcher generates a patch is  important for validity. We explored a few options here. Unified diff, which is the standard format for git patches, contains headers beginning with \texttt{@@} for each hunk, and added (deleted) lines prefixed with a \texttt{+} (\texttt{-}). This is a very rigid format that is unnatural for LLMs, given that a large part of their training is on raw code without every line being prefixed with a + or -. We also explored line diff, which represents a patch compactly by only listing the added and deleted lines prefixed with line numbers.

The final patch format is based on the search-replace format proposed.\footnote{Search-replace diff format \url{https://aider.chat/docs/more/edit-formats.html}} This intuitive format consists of a ``search'' block, representing the code to search for, and a corresponding ``replace'' block to replace it with. The only constraint is for the search block to be unique in a given code context, as if there are multiple instances of it, there is no way to know which one to pick for replacing. Therefore, we prompt the patcher to produce search-replace blocks with these conditions.


Below we show an example of the prompt provided to the patcher:

\begin{tcolorbox}[title={\textbf{Patching: Prompt}}] \label{sys_prompt}
You are a proficient programmer assisting a colleague with code updates.
You'll be given the code and a description of the required changes.
Contextual information will be provided to support your task.
Consider the most effective methods to edit the code. You must respond in the format shown below.

$<$$<$few shot examples$>$$>$
\\ \\
\texttt{[code]}

\texttt{path/to/file.py}

\textit{$<$code\_in\_file.py$>$}

\texttt{[instruction]}

\textit{$<$the task-specific edit instruction from the orchestrator$>$} \\

Format your changes as \diffs using SEARCH/REPLACE block rules.
[... other formatting instructions...]
\end{tcolorbox}

\subsection{Patch Generation Benchmark}
\label{subsec:patchgen}

To isolate the ability to generate the final patch from the orchestration stages, we create a separate \patchgen benchmark. The \patchgen benchmark is of the form \emph{$<$Input file, Natural language instructions, Test oracle$>$}. Particularly, we take a test failure task and the ground truth \diff which fixed the task. From this we extract the file (\emph{Input file}) which was modified to fix the test. This step ensures localization of the task to the file.

To generate the natural language instructions, we provide the fix diff to a Llama model.  We use the ``backtranslation'' technique to generate natural language instructions from the human written patch that fixed the failing test. These NL backtranslation of the code fix are similar to the steps and instructions that someone would provide to fix the patch. 

To understand the impact of instructions, we backtranslate two variants: high-level instructions vs detailed instructions. This helps us understand (a) if the model can follow a complicated list of instructions (b) can the model generate the patch if we spoon feed it? This somewhat removes the ``reasoning'' aspect from the problem. By using high-level and detailed instruction, we can estimate a lower bound and upper bound of the patching system's performance. 

To illustrate with an example, we provided a diff and the backtranslated instructions (high-level and detailed) below.

\begin{tcolorbox}[title={\textbf{Back translation: high-level and detailed instructions}}] \label{back_translation}
\textbf{Diff}\\
\texttt{[code]}
\textit{Logger.warn("Logging the failures in function X")}\\

\textbf{High-level instructions}\\
\texttt{[NL]} \texttt{``can you add logging to function X?''}\\
\\
\textbf{Detailed instructions}\\
\texttt{[NL]} \texttt{``1. import logger''}\\
\texttt{[NL]} \texttt{``2. Add a log statement after the loop in X''}\\
\texttt{[NL]} \texttt{``3. Add a log message ``Logging the failures in function X''''}
\end{tcolorbox}

In total, we curate 210 tasks for \patchgen, consisting of 70 tasks each from three categories of patch complexity: small ($<$ 10 lines), medium (10-40 lines), large ($>$ 40 lines). The final oracle for \patchgen is whether the original test that failed now passes with the patch generated from the NL instructions on the input file.

\section{Validation}
\label{sec:ValidationMethod}
One of the biggest challenges with LLMs is they do not know what they do not know. This can lead to them generating suboptimal code that is composed of lint errors, build errors, and even test failures. 
Tools such as static analyzers and linters provide useful feedback to the agent the same way they aid human engineers to maintain correctness and quality guarantees. Additionally, these tools can enforce nonfunctional requirements such as styling, documentation, and reliability. Primarily, we employ two classes of deterministic tools: static analysis tools and test execution tools.

We combine neural and symbolic methods \cite{10.1145/3563327} to enforce quality constraints on the patches generated by LLMs. We run these validity checks at two places in the agentic workflow in Figure~\ref{fig:pipeline}: 1. inside the ReAct loop 2. at the end of the ReAct loop. These symbolic checks help validate whether the agent-generated solution is accomplishing the task at hand with soundness guarantees.

To this end, we employ the \agent with deterministic programming language tools that can enforce quality control during the patch generation (inside the ReAct loop) and post patch generation. We give the \agent access to validators in a similar fashion to popular integrated development environments (IDEs) that human software engineers use, as immediate feedback after making an edit or saving a set of edits. After an edit observation (tool result from performing an edit) with feedback from validators, we insert the following prompt: 

\begin{tcolorbox}[title={\textbf{\Diagnostics: Feedback Example}}] \label{sys_prompt}
The edit was successful but $<$name of static analyzer$>$ detected errors. Please review the changes and ensure they are correct and that additional errors were not introduced by this edit. You are able to edit the file again if necessary.

[...static analysis output...]
\end{tcolorbox}

\subsection{Test Execution Tools}

Similar to \diagnostics, we hypothesize that supplying the \agent with feedback from test execution will improve the quality of its work. Primarily, our goal for the \agent is to fix a failing test. Specifically, we hope that test execution feedback will give the \agent the right information (stack traces) to root cause and fix the failing test.

When a human software engineer undertakes a program repair task, their first goal is to reproduce the failure. Once the failure has been reproduced, an engineer would try to root cause the failure then determine a suitable solution and change the code. After making changes the engineer would again try to reproduce the failure, then continue to make changes until the failure no longer occurs. For program repair, test execution feedback is the mechanism by which we determine that the goal has been met.

In our implementation, we gave the \agent access to a test execution tool which runs the oracle tests on the intermediate agent-generated commits. Then, it offers the test execution feedback (test output, stack traces) as an observation. If the test passed, then the output is simply "All tests passed" but if the test failed then details of the failure like an assertion message or exception stack trace are supplied (see the example of a stacktrace prompt in Section~\ref{harness}). 

\subsection{LLM-as-a-Judge}
\label{secJudgeMethod}
A patch generated in response to fixing a test failure may fix the failing test but it might not be aligned with human preferences. For example, the patch could achieve functional correctness in a suboptimal manner, by using legacy libraries, or even induce additional stylistic changes human engineers do not like to see in their code. 

To address this problem, we leverage an LLM to act as a judge to help predict the likelihood of a patch matching human preferences. To that end, we calibrate a smaller iCodeLlama model~\cite{CodeCompose} (similar to the one used for backtranslation), with a many-shot prompt. The structure of the prompt is described below.

\textbf{Outcome: Binary Classification} The prompt instructs the LLM to classify the input patch into one of two classes:
        \begin{itemize}[label=\textbullet,leftmargin=0.2in]
            \item Class 0: The code is likely to be \textit{unacceptable} to a human engineer
            \item Class 1: The code patch is likely to be \textit{acceptable} to a human engineer
        \end{itemize}

\textbf{Prompt} Consists of instructions to perform the binary classification task and ten examples for each class. These examples are carefully crafted to reflect variants of acceptable and unacceptable patches. \\

\textbf{Input} Code patch (generated by the \agent) \\

\textbf{Output} The LLM is instructed to respond with the following:
        \begin{itemize}[label=\textbullet,leftmargin=0.2in]
            \item Reason: A natural language explanation of why the LLM classifies the patch as acceptable or unacceptable.
            \item Class: An integer representing the classification.
        \end{itemize}

Diffs that are classified as \textit{acceptable} (Class 1) are passed further in the pipeline (for code reviewing). Diff that are classified as \textit{unacceptable} (Class 0) are discarded and that information is logged for suture analyses and LLM calibrations.

\textbf{LLM-as-a-Judge benchmark.}
To evaluate the efficacy of the LLM-as-a-Judge approach, we created a benchmark of 244 data points. We asked human engineers to review these 244 \agent-generated patches and classify them into classes: \textit{acceptable}, \textit{unacceptable}. Human engineers labeled 57 as acceptable and 187 as unacceptable. Then, we tested our pre-calibrated LLM on this data set to calculate metrics such as precision and recall.

\section{Evaluation Methodologies and Context} 
\label{evaluation-method}

\subsection{Company Context}

\Meta is a large online services company that works across the spectrum in the communications, retail, and entertainment industries. Engineers at \company work in a rolling environment where the code base is constantly changing. Much of \company's engineering tools are built internally. Importantly, internal ownership and development of engineering tools provides access to internal usage data that can be leveraged to build benchmarks and measure online experiments~\cite{CodeCompose, Dunay2024FSE-industry, SQLCompose}. 
In this paper, we focus on the specific test failure workflow that we described in Section~\ref{secTestFailureBot}.

\subsection{Offline Evaluation Methodology}

To evaluate the \agent's ability in accomplishing program repair tasks, we ran them against the \benchmark. The benchmarking data points are in the form of a triplet $\langle Specification, Patch, Oracle \rangle$.
We provide the \agent with an input (Specification) and expect it to generate code (Patch) to accomplish the task mentioned in that specification. Then, we run the same tests that are broken (and flagged by the \testfailurebot) to check if the agent-generated patch is fixing them (Oracle).
We test the \agent's capability to solve the failed tests in four different settings. We ablate the experiments by passing different types of feedback from static analyses tools, test execution, or even a combination of both to check which setting helps the agent the most towards solving the test failures. The agent can either leverage the feedback generated by running the Oracles or can choose to ignore it.

\begin{enumerate}
    \item \textbf{ReAct agent} in this setting, we let the \agent approach the problem of solving a test failure using the Agentic harness (as explained in Section \ref{harness}). We do not pass any intermediate feedback generated by executing the Oracles (test cases) to the agent while the agent is attempting to solve the problem.
    \item \textbf{ReAct agent + \diagnostics Feedback} in this setting, we let the agent have access to the intermediate feedback generated by running static analysis tools (\diagnostics) while generating a patch. This helps the agent course-correct by fixing the \diagnostics errors or warnings in iterative fashion.
    \item \textbf{ReAct agent + Test Execution Feedback} in this setting, we ablate the experiment by passing the Test Execution Feedback (but not the \diagnostics feedback).
    \item \textbf{ReAct agent + \diagnostics Feedback + Test Execution Feedback} in this setting, we pass the feedback from \diagnostics and test execution and let the agent act on that feedback while solving the test failure tasks.
\end{enumerate}

Additionally, we run the \agent repeatedly by setting the LLM temperature to 0.8 \cite{brown2024largelanguagemonkeysscaling} and collect \textit{pass@10} results. Since re-running the test is automatic, we consider a solution to be correct if any of the 10 solutions generated by the agent fixes a test failure.

With this setup in place, we run the Agent on the \testfailurebot benchmark and measure the following metrics:
\begin{enumerate}
    \item \textbf{Solve Rate (SR)} is a metric that checks if the Oracles (unit tests that were failing) are passing now.
    \item \textbf{Patch Generation Rate (PGR)} is a metric that checks if the \agent is able to generate a patch even if it cannot be applied to the existing codebase. We noticed cases where agent-generated patches cannot even be applied to the program because of line number mismatches, premature termination of trajectories, etc.
    \item \textbf{Iteration Count (IC)} is a metric that measures the number of steps or iterations the \agent took to accomplish a task. These steps typically include actions such as using a tool or generating an intermediate patch. The larger the number of iterations the more expensive the solution.
    \item \textbf{Error Rate (ER)} is a count that measures whether errors occurred during the trajectory of the agent solving a task.
\end{enumerate}

\subsection{Evaluation Methodology for in Production} 
\label{evaluation-production}
\label{qualitative_feedback}
\label{online_metrics}

After evaluating the agent's performance on our offline benchmarks (Section \ref{evaluation-method}), we progressively rolled out the agent to run on \{10\%, 50\%, 100\%\} of production test-failure tasks.
During initial rollout, aside from the automatic validation (Section \ref{Validation}), we also randomly sampled the generated \diffs and had our team manually annotate whether the agent had correctly solved the task. This ensured a high quality bar for initial production release.

\textbf{Online Metrics.} We monitored the following metrics to evaluate the agent's performance on production tasks:

\begin{itemize}
    \item \textbf{Published Volume}: \textit{\# \diffs Published.} Represents the potential impact of the agent, with how many test failures it was able to generate a solution for. Published \diffs are then reviewed by human engineers.
    \item \textbf{Review Rate}: \textit{\# \diffs Engineers Reviewed / \# \diffs Published.} Tracks the online quality of the agent with how often engineers engage in reviewing the generated code. "Review" means an engineer commented, accepted, or requested changes.
    \item \textbf{Land Rate}: \textit{\# \diffs Landed / \# \diffs Published.} Tracks the online quality of the agent with how many \diffs successfully pass human review and are approved to land.
    \item \textbf{Landed Volume}: \textit{\# \diffs Landed.} Represents the final impact of the agent with how useful these \diffs were in automatically solving test failures, instead of engineers having to solve them.
\end{itemize}

\textbf{Qualitative Feedback.} We also aggregated feedback from comments left by engineers (1) on the agent-generated \diffs (2) in our engineer feedback group. This gave us a qualitative lens on whether the agent was solving tasks correctly, and whether engineers found the agent helpful. Using an open coding methodology, we reviewed and manually annotated a sample set of 100 comments. We identified recurring themes of usage and improvements that we rolled out to improve the \agent in production.


\section{Offline Results from Benchmarks}
\label{secOfflineResults}

In this section, we discuss the results for our offline benchmarks. We first discuss the ability of patching sub-agent to generate patches using different underlying models and diff formats. Based on the top sub-agent, we test the entire \agent on a benchmark. Finally, we report on the LLM-as-a-Judge that determines whether the generated patch conforms to the standards of engineering at \Meta.

\subsection{\patchgen Benchmark Results}
\label{subsec:ra4patching}

\begin{table}
\centering
\caption{Evaluation on the \patchgen Benchmark. We see that search-replace (Srch-Repl) is the most optimal diff format. This is especially true with detailed instructions with an improvement of 23pp over the unified diff format (for the 405B model). Comparing the two models, iCodeLlama-70B, that is internally fine-tuned for patch generation using the search-replace format, is highly competitive with the much larger but vanilla Llama-405B. We use the smaller specialized model and the search-replace format for the remainder of this paper.}
\begin{tabular}{l|r|r|r}
\hline
\textbf{Model}	&	\multicolumn{3}{c}{\textbf{SR for High-level / Detailed Instr.}} \\
\cline{2-4}
& \textbf{Unified diff} & \textbf{Line diff} & \textbf{Srch-Repl} \\
\hline
Llama-405B (public) & 26\% / 30\% & 20\% / 26\% & 42\% / 53\%  \\
iCodeLlama-70B (internal) & 16\% / 23\% & 16\% / 22\% & 43\% / 51\%  \\
\hline
\end{tabular}
\label{tab:PatchGenResults}
\end{table}

We compared the patcher performance on the \patchgen benchmark that contains 210 tuples: \emph{$<$Input file, NL instruction, Test oracle$>$} as described in Section~\ref{subsec:patchgen}. 
Since we want to isolate the effect of patch generation, we evaluated it in a non-agentic, zero-shot setting, by providing the model the input file to edit and the backtranslated natural language instruction extracted from the human patch fix. 
Our evaluation metric was the best Solve Rate (SR), \ie, whether the failing tests passed after applying the generated patch.

We have three sub-research questions:

First, what is the impact of diff format on solve rate? We evaluated three formats: (i) unified diff, (ii) line diff, (iii) search-replace. 

Second, what is the impact of high-level vs detailed instructions on solve rate? We vary the natural language instructions to contain either a high-level task or very specific solution instructions. Full details on the NL instructions can be found in Section~\ref{back_translation}.

Third, which Llama-3.1 based model has the highest solve rate? In the study we evaluate two models: (i) the large Llama-405B public, (ii) and a smaller internally fine-tuned Llama-70B model (iCodeLlama-70B)~\cite{CodeCompose,Dunay2024FSE-industry}.

\textbf{Diff Format Results.} In Table~\ref{tab:PatchGenResults}, on the same model, say Llama-405B, we see that the best solve rate for unified diff, line diff, and search-replace diff formats is 30\%, 26\%, and 53\%, respectively. 
The standard unified format used in git, is unnatural for LLMs, given that a large part of their training is on raw code without every line being prefixed with a + or -. While the line diff that only shows changed lines, works for small patches (2-3 lines), as the patches grew larger and spread across hunks, the patcher started struggling to keep track of the line numbers. This format does not contain enough context and actually has a 4pp lower solve rate than the unified diff format. The search-replace format has the highest solve rate, a 23pp and 27pp improvement over the other two formats, respectively. We believe that this format is very natural for LLMs to produce, as the code in both blocks is without any additional prefix or line numbers, much like the large volume of code the LLM was trained on. For the remainder of this paper, we use the search-replace diff format for our \agent.

\textbf{Highlevel vs Detailed Instruction Results.} 
In Section~\ref{back_translation}, we provide an example of a highlevel instruction vs the detailed step-by-step detailed instructions. As we would expect and see in Table~\ref{tab:PatchGenResults}, the detailed instructions provide an increase of 4 to 11 percentage points improvement in solve rate across the various patch formats and models.

\textbf{Llama-405B vs iCodeLlama-70B Model Results.} In Table~\ref{tab:PatchGenResults}, we observe that a custom fine tuned iCodeLlama-70B model, offers a competitive alternative compared to Llama-405B, a significantly larger foundation model. Particularly, on the search-replace format that it is fine-tuned on, it is even able to outperform the Llama-405B model when provided with high-level instructions. For optimizing between model performance and scalability, in the remainder of this paper we use iCodeLlama-70B as the patch generation model.

\begin{table}
\centering
\caption{Offline \agent Ablation Results (\textbf{bold} indicates the best result for each category). The offline benchmark shows the impact of the neural model, ReAct, with the addition of different types of symbolic information, \ie static analysis feedback and test execution information. The final model used in production balances solution quality with cost and latency: ReAct agent + Static Analysis + Test Execution @ 1.}
  \begin{tabular}{lrrrrrr}

\toprule

\textbf{Setup}	&	\textbf{SR}	&	\textbf{SR}	&	
{\textbf{PGR}} &	
{\textbf{ER}} &	
{\textbf{IC}}	\\ 
 & @ 1 & @ 5 & avg & avg & med \\ 
\hline

ReAct agent & 28.5\% & 46.3\% & \textbf{87.2\%}  & \textbf{0.0\%} & \textbf{6.8} \\ \\
ReAct agent \\ + Static Analysis & 34.1\% & 49.0\% & \textbf{87.2\%}  & \textbf{0.0\%} & 8.4 \\ \\
ReAct agent \\ + Test Execution & \textbf{43.9\%} & \textbf{61.0\%} & 87.0\%  & 1.0\% & 12.2 \\ \\
ReAct agent \\ + Static Analysis \\ + Test Execution & 42.3\% & 58.5\% & 85.9\%  & 0.2\% & 11.8 \\

\bottomrule

  \end{tabular}
  \label{tab:OfflineResults}
\end{table}

\subsection{Agent Benchmark Results}
\label{sec:AgentBenchmarkResults}

The \patchgen benchmark is artificial in that the natural language instructions are generated from the actual diff fix. We now turn to the actual usecase of generating a patch from the real, unseen test failures raised by \testfailurebot. In this offline evaluation, we perform the \agent loop shown in Figure~\ref{fig:pipeline} going from a test failure, through reasoning, and ending in validation. We constructed a benchmark to evaluate the \agent in Section~\ref{sec:agentBenchmark}. We perform ablations to understand how important the ReAct agent (neural model) is compared to the symbolic information from static analysis and test execution traces. We also report the number of iterations as well as the running the agent multiple times, up to 5, to see if different runs improve performance. 

The results are shown in Table~\ref{tab:OfflineResults}. Using only the {\it ReAct harness} the \agent has a solve rate of SR@1 of 28.5\%. We see the highest SR@1 rate is 43.9\% when the test execution feedback is passed to the \agent after each code edit.
When we add only {\it static analyses tools} the SR@1 goes from from 28.5\% to 34.1\%. This requires more iterations because the agent must re-run the ReAct loop when there are static analysis failures and the IC went up from 6.8 to 8.4, which incurs more computational cost per solution as the LLM needs to process more tokens. 

When we add only the {\it test execution feedback} we attain the best performance with SR@1 going from 28.5\% to 43.9\%. The IC goes from 6.8 to 12.2 which further increases the computational cost per solution as test execution feedback stack traces are much more verbose and consumes more tokens in the LLM context window. Additionally, executing tests at every iteration will also incur extra infrastructure cost and causes the \agent to take more time to produce a solution. 

When both the {\it static analyses and test execution feedback} are passed together, the Error Rate (ER) went down significantly compared to the test execution only from 1.0\% to 0.2\% with a reduction in SR@1 from 43.9\% to 42.3\%. Interestingly, even though there is more feedback being passed to the ReAct loop, the IC count actually goes down to from 12.2 to 11.8. This indicates that static analyses tools help the \agent recover from certain classes of errors that the test execution feedback alone cannot alleviate. 
 
To test whether {\it repeated runs} can have any impact on solve rate, we ran the \agent five times on all the tasks in the benchmark. Each run is independent and a full run of the \agent. We see the solve rate with five repeated runs, \textit{SR@5} is 61.0\% when only test execution feedback was passed to the \agent. This is a 17 percentage point improvement over a single run, SR@1 = 43.9\%. The improvement in solve rate is observed consistently across all the settings when sampling is applied. This is inline with the hypothesis that scaling inference compute with repeated sampling does have a positive impact on LLM's ability to solve tasks, at the expense of costing more resources and time \cite{brown2024largelanguagemonkeysscaling}.

We decided that {\it in production} a single run, SR@1, with ReAct and both the static analysis and test execution feedback was the right balance between solve rate and error rate and the computational costs and latency requirements.

\subsection{LLM-as-a-Judge Benchmark Results}
Although the \agent now has an impressive solve rate, we wanted to be sure about the quality of the patch and its alignment to human engineer preferences. To that end, we calibrate an LLM to act as a ``judge'' and create a benchmark to test the ability of an LLM to judge a patch's quality per human engineer preferences. The calibration process and the benchmark are explained in Section~\ref{secJudgeMethod}.

When we evaluated out LLM-as-a-Judge on the benchmark, the LLM judge identified 72 true negatives (identified `unacceptable' patches as `unacceptable') at the cost of 11 false negatives (identified `acceptable' patches as `unacceptable'). Our goal is to not ship more patches but to reduce noise as much as possible as shipping low-quality patches will waste human code review time and might even make the engineers lose trust in our system.

Therefore, we optimized for a higher Class-1 precision. As shown in the Table~\ref{tab:classwise_metrics}, after going through multiple calibration optimizations and iterations, we landed on a prompt that yielded a high precision for Class-0 .86 allowing us to use this judge in production to reduce the number of bad/unacceptable patches that are shown to engineers. 

\begin{table}
\centering
\caption{Precision and Recall for LLM as a Judge. We use the judge to reduce the number of bad AI fixes that are shown to engineers. We focus on high unacceptable/bad patches, class 0, precision. We attain a high precision of .86 which will reduce the number of inappropriate patches shown to engineers in production.}
\label{tab:classwise_metrics}
\begin{tabular}{l|r|r}
\hline
\textbf{Class} & \textbf{Precision} & \textbf{Recall} \\
\hline
Class 0 (Unacceptable) & $0.286$ & $0.807$ \\
\hline
Class 1 (Acceptable) & $0.867$ & $0.385$ \\
\hline
\end{tabular}
\end{table}

\section{The \agent in Production}
\label{secProduction}


The product experience follows the major stages outlined in Figure~\ref{fig:pipeline}: Test failure task, agentic harness and patch generation, validation, code review, and task closed. We discuss each below and show an example output for the agent's trajectory in Figure~\ref{fig:product-traj}.

\subsubsection*{Test failure task} When a broken test is detected by the \testfailurebot tool, a task is automatically generated and assigned to the responsible owner. 
Upon task creation, the \agent is automatically triggered in a background job to generate a fix \diff for the failing test. 

\subsubsection*{Agent harness and patch generation} The agent is given the task as its input specification, then walks through the ReAct loop (Section \ref{harness}) to reproduce and fix the broken test (the agent workflow is shown as a trajectory in Figure \ref{fig:product-traj}).
 
\subsubsection*{Validation} \label{Validation} Once the \agent generates a patch, we validate the changes in three steps:
\begin{enumerate}
    \item \textbf{Breaking Test}: Validate that the originally breaking test is now passing.
    \item \textbf{LLM as a Judge}: We trained a Llama-based judge to determine if the patch is of reasonable quality, see Section~\ref{secJudgeMethod}.
    \item \textbf{Continuous Integration}: The generated patch is submitted to a continuous integration system that automatically builds, tests, and validates changes across relevant parts of the codebase. 
\end{enumerate}

After the above validation signals all pass, the repair patch (diff) is automatically submitted for review.

\subsubsection*{Code Review}
\begin{figure*}
    \centering
    \includegraphics[width=.8\linewidth]{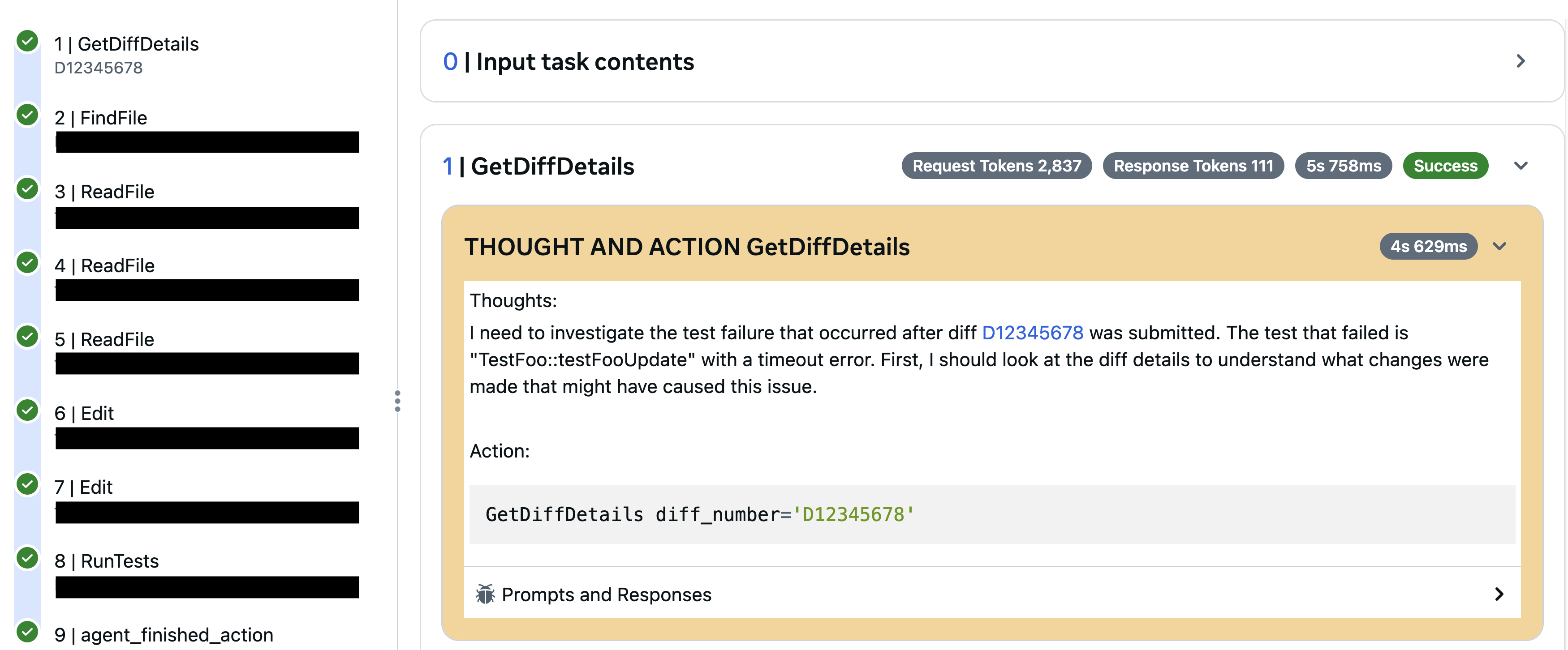}
    \caption{An example of the trajectory of the agent fixing a broken test. This information is also provided to the reviewer so they can evaluate the intermediate steps to generate the patch fix.}
    \label{fig:product-traj}
\end{figure*}

To facilitate the review of AI generated code, we provide the following in the submitted diff:
\begin{enumerate}
    \item A link to the agent trajectory (Figure \ref{fig:product-traj}) that created this diff
    \item Validation signals from \ref{Validation}, such as the passing test results and CI signals, are included to increase confidence in the generated code 
    \item The associated \testfailurebot task is attached in the \diff 
\end{enumerate}

These \diffs go through the standard review process by humans, directed to be reviewed by the appropriate teams which own the breaking tests.

\subsubsection*{Task Closed}
Once the \diff is accepted and landed, we verify the test failure is resolved on the main branch, then automatically close the associated task. 

\subsection{Production Usage Results}

\begin{table}
\centering
\caption{Online Production Results (Feb 1 - Apr 30, 2025). 80\% of the AI fixes were of sufficient interest/quality that engineers took time to review them. Of those that received a review, 31.5\% were accepted. }
\label{tab:online_metrics_table}
\begin{tabular}{l|r}
\hline
\textbf{Item} & \textbf{3 Month Total} \\
\hline
\textbf{\# \diffs Published} & 1589 \\
\hline
\textbf{\# \diffs Reviewed} & (80\%) 1285 \\
\hline 
\textbf{\# \diffs Landed} & 405\\
\hline
\textbf{Land Rate of Total} & 25.5\%\\
\hline
\textbf{Land Rate of Reviewed} & 31.5\%\\
\hline
\end{tabular}
\end{table}

We deployed the \agent to solve live test-failure tasks normally assigned to engineers across \company. We tracked the metrics outlined in Section \ref{online_metrics} to monitor the online performance and impact. Over a three-month period, the \agent published nearly 1.6k \diffs, with 25.5\% of these accepted and landed by human engineers who deemed them to be of sufficient quality to correctly solve the broken test (Table \ref{tab:online_metrics_table}).

Prior works reported the acceptance rate for AI-assisted code completions in an IDE at 22~\cite{CodeCompose} and 21~\cite{SQLCompose}. In this work, the \agent is suggesting an entire solution to a broken test which is a much more difficult problem. We see a total acceptance rate of 25.5\%, indicating that our \agent not only generates good solutions, but that our validation loop and LLM judge catch low quality solutions before they are shown to engineers.

\subsection{Developer Feedback Results}

Among \diffs with comments, we monitored qualitative  feedback to ensure the quality of the agent in solving production test failure tasks. By manually annotating a sample set of 100 comments (Section \ref{qualitative_feedback}), we identified the following recurring themes of feedback. Below, we explain each theme along with grounding examples, followed by classes of improvements we made to the agent to address the feedback.

We first discuss some positive sentiment from engineers, we later discuss negative feedback and fixes we rolled out. Some positive themes are below (n=8):

    \textit{Quick approval}: They would accept the \diff with a brief comment "stamping" the change, in the same way that they would quickly accept a straightforward, valid change from another human on the team. 
 
    \textit{Gratitude}: Engineers would explicitly thank the agent for doing its work, particularly when it demonstrated advanced capabilities fixing a test deemed too complicated to be solved without human intervention. Conversely, even for easy fixes, they would be grateful that the agent saved them time.
    
    \textit{Surprise}: They were pleasantly surprised in cases where the agent fixed tests they did not even know were failing for their oncall. Some agent fixes exposed gaps in the underlying continuous integration system that allowed the original blame change to pass through.

Below are some representative quotes:

\textit{"I got excited, checked out the code, debugged a bit, only to realize it was already fixed in trunk by the \agent in under 24 hours (I'm guessing a good chunk of that was code review + merging time). Cool to see where things are at these days."} 

\textit{"Wow I thought the LLM was being dumb in commenting out the test check, but it's actually precisely right since the blame D\#\#\#\# commented out the function call while trying to deal with the SEV} 

\textit{"Amazing; instead of reverting this \diff, proposes just fixing the core logic."}

\textbf{Test already fixed (n=47).} A significant number of review comments fell under the umbrella of the test already being fixed when the agent published its \diff. Another factor was flaky tests. 

Example comments:

    \textit{"This is almost certainly a flaky test rather than a problem with this pattern.  I don't know where the agent came up with this idea, but I'd rather look at the flakiness of this test instead."} 

We addressed these testing issues with the following improvements:

    \textit{Test Validation}: We added an additional test run to validate that the test was reliably failing, *before* the agent would even start its ReAct loop. This way we would have a deterministic signal on test flakiness.
    
    \textit{Infra reliability}: We moved the agent to run in a container with a more consistent environment to increase the chance of the test reliably reproducing.
    
    \textit{Rules Engine}: We added simple rules to check if human engineers had already produced a fix. For example, if the original task had any attached \diffs, we would not publish the agent-generated change.

\textbf{Difficulty finding reviewers (n=6).} Some reviewers would comment asking other teammates, or even engineers from different orgs, to take a look -- as they themselves were not familiar with the code change. Consequently, some agent changes would sit in queue without ever being reviewed. For example:

\textit{"I have no context on this so probably can't review it. @colleague1 are you okay finding someone suitable?"}

To \textit{improve} the review rate of the agent's changes, we worked with other engineering teams to set up incentives for reviewing more AI-generated code. We also adjusted the title and description on the \diff to make it easier to understand the agent's reasoning process, so it would be more transparent to review.    

\textbf{Missing tools (n=3).} Several engineers noticed cases where the agent attempted to publish \diffs despite failures in linters, formatters, and code generators across various platforms. For example, \textit{"I would love to know how the \diff that triggered this which made GraphQL related changes, then caused a bot to look at and modify a totally unrelated project (which doesn't use GraphQL), auto-generating a change that broke our tests for our simulator CI tool, which it claims were rerun to confirm everything passed"}


From this feedback, we identified several missing debugging tools and validators that we could add into the agent's ReAct loop so that the agent could run these signals earlier and fix the failures.

\textbf{Partially correct solutions are useful (n=19)}. While our benchmarks evaluated the agent in binary fashion on whether oracle tests passed or not (Section \ref{evaluation-method}), we learned that in a real-life production setting, humans found partial solutions valuable and would often build on top of the agent's work for several reasons:

    \textit{Provides Starting Point}: Seeing the agent's code change helped some engineers identify exactly where the error was happening e.g. localizing the issue to the correct file so humans did not have to start from scratch to write the correct final solution.

    \textit{Saves Time}: Even when the agent fix was not the ideal solution, if it improved on the current state, human engineers would still land the change to incrementally improve the codebase.
    
    \textit{Initiates Discussion}: Several teams found it helpful that the agent's \diff triggered a discussion on the right approach to fix these tests, as the proposed changes often revealed some deeper complexity in the stack that the team had not been aware of

The following is a grounded example of the theme: \textit{"This is pretty neat! The fix is in the right direction but instead of an Ok() || PartialSuccess() we should just expect the status to be PartialSuccess() in all of these cases that were changed. It's a bit more nuanced and needs case by case analysis. I can take care of the fix. But I'm excited to see the agent work here !"} 
    
    



\textbf{Identifying source vs. test changes (n=6)}. For our benchmark, we only allowed data points where the human solution changed source files, and not test files. However, in production, there is no way a priori to know whether a failing test is correctly fixed by source file vs. test file changes. Indeed, we encountered frequent feedback that the agent either reverted the blamed change or modified source files -- instead of "fixing forward" by changing the test file to match the source file changes that caused the test to fail.

    \textit{D\#\#\#\#: the right fix is to fix the typo in the test instead, though testing getDescription doesn't make that much sense to begin with}  
    

To adjust for the need to potentially modify tests in production test failures, we did the following:

    \textit{Revert Validator}: We added a validator that would check how closely the agent code changes reverted the original blame change. Highly similar changes would give a failing signal.
    
    \textit{Test Flag}: We enabled a flag allowing the agent to selectively change test files to experiment with its impact on land rates.

\textbf{Failing to understand production environment (n=11)}. Even if the agent's code changes made sense in isolation to solve the broken test, it would lack context on the impact of certain features rolling out in production:

 
    \textit{"LLM feedback? this would have removed a flag that's actually in use in the TTLS binary.}

\textbf{Conclusion.} Autonomously solving test failures at scale across a diverse set of production code bases at \company poses a wide range of difficulties, as shown in the various categories of engineer feedback. A given test failure may have many possible causes and solutions that differ in test vs production environments. It is essential to provide the agent with the right context and tools, along with a full suite of validation techniques to generate the highest quality \diffs for human engineers to review and land.

\section{Threats to Validity}
\label{sec:threats}

\subsection{Generalizability}

Drawing general conclusions from empirical studies in software engineering is difficult
because any process depends to a large degree on a potentially large number of relevant
context variables. For this reason, we cannot assume a priori that the results of a study
generalize beyond the specific environment in which it was conducted~\cite{Basili1999TSE}. The processes, environment, and culture of a company may influence the results of a study. Therefore, it is unclear whether our findings can be generalized to other contexts. 

%


\subsection{Construct Validity}

One potential construct validity threat to our study is the evaluation metrics we used to assess the models' performance offline and in production. We measured the solve rate, patch generation rate, the iteration count, and the error rate for offline models. A passing test and a successfully generated patch does not always mean that the patch will fix the code correctly. As a result, we also measured the published volume, reviewer rate, and the land rate in production. While the land rate indicates how many generated patches were actually merged and pushed, there was a large number, 20\%, that never received review, so it is possible that if we had enough reviewers we would have seen even higher land rates. 

\subsection{Internal Validity}

Our offline studies used benchmarks validated by our engineers: 210, 123, and 244 for the \patchgen, \agent, and LLM-as-a-Judge respectively. While the benchmarks are small they are on par with those used in other agentic program repair~\cite{GoogleAgentProgramRepair}. 
The offline results served only as a starting point to ensure that the models would be reasonable in production. We used a progressive random rollout to 10\%, 50\%, and finally 100\% of test failures. 

The qualitative feedback was gather from the feedback group for \agent that all employees have access to. The feedback is not a random sample and could have bias. The feedback is publicly available for other employees to see. However, prior works have shown that this type of feedback usually receives more negative feedback and suggestions for improvements than would be expected by chance~\cite{CodeCompose,SQLCompose}. Future research could involve surveys of a random sample of engineers. We were able to triangulate some of these findings based on the number of \agent patches that are reviewed and landed.

\section{Literature and Discussion}
\label{sec:literatureAndDiscussion}
The APR community has shown considerable enthusiasm for autonomous workflows. However, the proposed approaches in this area have primarily been designed and assessed using open-source bugs from the GitHub ecosystem. As an example, SWE-Bench~\cite{jimenez2024swebenchlanguagemodelsresolve} is a benchmark sourced from GitHub issues across $12$ popular Open Source projects, which contains $2,294$ Python bugs and fixes. There is also a smaller version, SWE-Bench-Lite, with $300$ bug/fix pairs. SWE-Bench has emerged as the standard evaluation benchmark for APR. The benchmarks produced for \agent are sourced in a similar way from \company repository and tasks. SWEBench differs in that it is static vs. \company which is rolling. SWE-bench Verified~\cite{swebenchverified} is a human annotated subset of SWE-bench, that is deemed solvable and further composed into an ``easy'' and a ``hard'' set. Across various agentic harnesses OpenAI observed increased solve rates on SWE-bench Verified. 

Recently, addressing the complexity of creating benchmarks, SWE-smith has been proposed as a pipeline for generating software engineering tasks at scale~\cite{yang2025swesmithscalingdatasoftware}. This approach allows for the automatic synthesis of thousands of task instances that break existing tests in a given codebase --- hence, facilitating more comprehensive evaluations of APR methods. By addressing the limitations associated with manual benchmark curation, SWE-smith can help drive the development of more robust and scalable APR solutions. As opposed to the benchmarks produced by us at \company, there are no guarantees that the created tasks by SWE-smith are realistic. Another well-known benchmark is SWE-Lancer~\cite{miserendino2025swe}, which consists of over $1,400$ freelance software engineering tasks. Several other datasets exist, see~\cite{ferrag2025llm} for a more comprehensive listing.  

Rondon \textit{et al.}, from Google, present their approach to automated program repair, Passerine, by establishing a benchmark similar to SWEBench and a solver approach similar to SWEAgent~\cite{rondon2025evaluatingagentbasedprogramrepair}. They report that machine generated bugs are 3x as likely to be fixed by their system. This is similar to our \agent whose input are machine generated tasks. Google's benchmark is a static one, and they support ``rewinding'' the repository for evaluation. \company is rolling or evergreen. Unlike our \agent, in use at \company, Google's system is not productionized yet but rather focuses on offline benchmark evaluations.

Our work follows the initial large scale deployment at Google~\cite{rondon2025evaluatingagentbasedprogramrepair}. Google’s “machine generated tasks'' benchmark comprises 2 types of tasks: viz, TOD (Test Order Dependence) and SAN (Automatic Sanitizers). Meta’s ``machine generated tasks'' benchmark comprises Test Failure Bot tasks only. The work is a step-function over the Google work as a Test Failure Bot task could be caused by anything -- a dependency change, a bug in the code, a bug in the test, perhaps even flakiness. And, indeed, it can be ``fixed'' in various ways --- change the code, change a dependency, change the test, etc. The agent needs to identify and select between these options. Meanwhile, a TOD task specifically names two tests, and poses the problem ``these tests use shared state''. It would typically be clear that the fix is either to eliminate the shared state or to make sure the second-named test resets it. It is fully localised as an issue, and of a very constrained form.

RepairAgent~\cite{bouzenia2024repairagent} employs an agent with actions governed by a state machine, in order to restrict certain actions. AutoCodeRover~\cite{zhang2024autocoderoverautonomousprogramimprovement}, on the other hand, is an agent-based repair system that leverages explicit program information, such as class and method definitions, as well as test-based localization. Unlike our \agent, which implements a ReAct-style loop~\cite{yao2023react}, it uses a two step process: Context Retrieval and Patch Generation. AutoCodeRover's fixing capabilities have been evaluated on Github issues sourced from SWEBench.

Building upon AutoCodeRover, SpecRover~\cite{ruan2024specrover} introduces a natural-language specification for expected behavior at each candidate repair location and incorporates a patch review agent. CodeR~\cite{chen2024coder} decomposes APR into multiple sub-tasks, coordinated by a task graph created and reviewed by a ``manager'' agent. MarsCode Agent~\cite{liu2024marscode} combines a dynamic, iterative approach to program repair with a traditional generate-and-validate pipeline in a multi-agent repair framework. OpenDevin~\cite{wang2024opendevin} (now OpenHands~\cite{wang2025openhandsopenplatformai}) provides a flexible foundation for building agent-based solutions for various software engineering tasks and domains. 

SWEAgent~\cite{yang2024sweagentagentcomputerinterfacesenable} is a system built on a ReAct-style loop, utilizing an Agent Computer Interface (ACI) to provide the agent with access to various tools. Notably, these tools are command line oriented and grant the agent terminal access. In turn, it is worth noting that the \agent proposed in this paper does not currently possess direct terminal access.

Several approaches have been proposed to leverage Large Language Models (LLMs) for automated program repair, which have been shown to outperform all existing approaches~\cite{xia2023automated}. AlphaRepair~\cite{xia2022less}, for instance, is a zero-shot program repair technique that relies on LLMs to fix bugs without requiring any additional training data. FitRepair~\cite{xia2023automated} improves on AlphaRepair and incorporates the \textit{plastic surgery hypothesis}~\cite{barr2014plastic}, which suggests that the code necessary to fix a bug often already exists within the same project, to repair the program. ChatRepair~\cite{xia2024automated} takes a conversational approach to repair, engaging in a dialogue with the engineer to identify and correct errors. As opposed to these previous approaches, Agentless~\cite{xia2024agentless} adopts, as its name suggests, an agentless strategy, that avoids complex tools and decision-making processes in favor of a simple three-phase pipeline consisting of localization, repair, and patch validation. Agentless has been shown to achieve state-of-the-art performance at a remarkably low cost, outperforming existing open-source software agents.
\section{Concluding Remarks}
\label{sec:conclusion}
In this paper, we seek to understand the viability of performing agentic program repair at large scale. We chose to tackle the problem of fixing source code based on test failures at scale, across \company's diverse software offerings, as means to validate our hypothesis.

To that end, we built a RaAct-based agentic harness powered by a Llama model as base. We provided the agent with an environment to operate in and a comprehensive set of tools to perform various actions in that environment ranging from checking out source code to running builds and tests autonomously.

To validate the efficacy of \agent in solving test failures, we created a benchmark consists of 123 real test failures across 15 programming languages. Similarly, we curated other auxiliary benchmark to test the ability of the system to generate patches and judge bad patches. 

We ran various ablations by varying the base models, patch formats, validation feedback passed to the agent, etc. The balanced model yielded a solve rate of 42.3\% with an average number of 11.8 iterations.

Inspired by these findings we set out to test the \agent in production. In a three month period, 80\% of the generated fixes were reviewed, and 25.5\% of the total number of generated fixes were landed in production. Qualitative feedback from the engineers at \company has been positive and manifested in the form of quick reviews, gratitude, and even surprise.

In the future, we anticipate the proposed methodology, architecture, and the system we built can be scaled to solve various problems in Software Development Life Cycle (SDLC) such as modernizing code bases, reducing technical debt, improving code quality, synthesizing tests, or even optimizing performance of the source code.

\section{Acknowledgements}
We would like to thank Andy Chiu, Arun Ganesan, Hannes Verlinde, Yiru Zhu, Shahin Sefati, Zach Rait, David Recordon, and Kwaku Akoi for their help and support with this work.

\bibliographystyle{IEEEtran}
\bibliography{main}

\begin{thebibliography}{10}
\providecommand{\url}[1]{#1}
\csname url@samestyle\endcsname
\providecommand{\newblock}{\relax}
\providecommand{\bibinfo}[2]{#2}
\providecommand{\BIBentrySTDinterwordspacing}{\spaceskip=0pt\relax}
\providecommand{\BIBentryALTinterwordstretchfactor}{4}
\providecommand{\BIBentryALTinterwordspacing}{\spaceskip=\fontdimen2\font plus
\BIBentryALTinterwordstretchfactor\fontdimen3\font minus \fontdimen4\font\relax}
\providecommand{\BIBforeignlanguage}[2]{{%
\expandafter\ifx\csname l@#1\endcsname\relax
\typeout{** WARNING: IEEEtran.bst: No hyphenation pattern has been}%
\typeout{** loaded for the language `#1'. Using the pattern for}%
\typeout{** the default language instead.}%
\else
\language=\csname l@#1\endcsname
\fi
#2}}
\providecommand{\BIBdecl}{\relax}
\BIBdecl

\bibitem{Koyuncu_2020}
\BIBentryALTinterwordspacing
A.~Koyuncu, K.~Liu, T.~F. Bissyandé, D.~Kim, J.~Klein, M.~Monperrus, and Y.~Le~Traon, ``Fixminer: Mining relevant fix patterns for automated program repair,'' \emph{Empirical Software Engineering}, vol.~25, no.~3, p. 1980–2024, Mar. 2020. [Online]. Available: \url{http://dx.doi.org/10.1007/s10664-019-09780-z}
\BIBentrySTDinterwordspacing

\bibitem{huang2023surveyautomatedprogramrepair}
\BIBentryALTinterwordspacing
K.~Huang, Z.~Xu, S.~Yang, H.~Sun, X.~Li, Z.~Yan, and Y.~Zhang, ``A survey on automated program repair techniques,'' 2023. [Online]. Available: \url{https://arxiv.org/abs/2303.18184}
\BIBentrySTDinterwordspacing

\bibitem{10.1145/3318162}
\BIBentryALTinterwordspacing
C.~Le~Goues, M.~Pradel, and A.~Roychoudhury, ``Automated program repair,'' \emph{Commun. ACM}, vol.~62, no.~12, p. 56–65, Nov. 2019. [Online]. Available: \url{https://doi.org/10.1145/3318162}
\BIBentrySTDinterwordspacing

\bibitem{zhang2024systematicliteraturereviewlarge}
\BIBentryALTinterwordspacing
Q.~Zhang, C.~Fang, Y.~Xie, Y.~Ma, W.~Sun, Y.~Yang, and Z.~Chen, ``A systematic literature review on large language models for automated program repair,'' 2024. [Online]. Available: \url{https://arxiv.org/abs/2405.01466}
\BIBentrySTDinterwordspacing

\bibitem{7816488}
X.-B.~D. Le, Q.~L. Le, D.~Lo, and C.~Le~Goues, ``Enhancing automated program repair with deductive verification,'' in \emph{2016 IEEE International Conference on Software Maintenance and Evolution (ICSME)}, 2016, pp. 428--432.

\bibitem{10.1007/978-3-030-11245-5_4}
T.-T. Nguyen, Q.-T. Ta, and W.-N. Chin, ``Automatic program repair using formal verification and expression templates,'' in \emph{Verification, Model Checking, and Abstract Interpretation}, C.~Enea and R.~Piskac, Eds.\hskip 1em plus 0.5em minus 0.4em\relax Cham: Springer International Publishing, 2019, pp. 70--91.

\bibitem{Frenkel2022}
\BIBentryALTinterwordspacing
H.~Frenkel, O.~Grumberg, B.-C. Rothenberg, and S.~Sheinvald, \emph{Automated Program Repair Using Formal Verification Techniques}.\hskip 1em plus 0.5em minus 0.4em\relax Cham: Springer Nature Switzerland, 2022, pp. 511--534. [Online]. Available: \url{https://doi.org/10.1007/978-3-031-22337-2_25}
\BIBentrySTDinterwordspacing

\bibitem{rondon2025evaluatingagentbasedprogramrepair}
\BIBentryALTinterwordspacing
P.~Rondon, R.~Wei, J.~Cambronero, J.~Cito, A.~Sun, S.~Sanyam, M.~Tufano, and S.~Chandra, ``Evaluating agent-based program repair at google,'' 2025. [Online]. Available: \url{https://arxiv.org/abs/2501.07531}
\BIBentrySTDinterwordspacing

\bibitem{cheng2025agenticbugreproductioneffective}
\BIBentryALTinterwordspacing
R.~Cheng, M.~Tufano, J.~Cito, J.~Cambronero, P.~Rondon, R.~Wei, A.~Sun, and S.~Chandra, ``Agentic bug reproduction for effective automated program repair at google,'' 2025. [Online]. Available: \url{https://arxiv.org/abs/2502.01821}
\BIBentrySTDinterwordspacing

\bibitem{222607}
\BIBentryALTinterwordspacing
R.~Bhagwan, R.~Kumar, C.~S. Maddila, and A.~A. Philip, ``Orca: Differential bug localization in {Large-Scale} services,'' in \emph{13th USENIX Symposium on Operating Systems Design and Implementation (OSDI 18)}.\hskip 1em plus 0.5em minus 0.4em\relax Carlsbad, CA: USENIX Association, Oct. 2018, pp. 493--509. [Online]. Available: \url{https://www.usenix.org/conference/osdi18/presentation/bhagwan}
\BIBentrySTDinterwordspacing

\bibitem{yao2023react}
S.~Yao, J.~Zhao, D.~Yu, N.~Du, I.~Shafran, K.~Narasimhan, and Y.~Cao, ``React: synergizing reasoning and acting in language models (2022),'' \emph{arXiv preprint arXiv:2210.03629}, 2023.

\bibitem{GoogleAgentProgramRepair}
\BIBentryALTinterwordspacing
``Evaluating agent-based program repair at google,'' 2025. [Online]. Available: \url{https://arxiv.org/pdf/2501.07531}
\BIBentrySTDinterwordspacing

\bibitem{ProbabalisticFlakiness}
\BIBentryALTinterwordspacing
``How do you test your tests?'' 2020. [Online]. Available: \url{https://engineering.fb.com/2020/12/10/developer-tools/probabilistic-flakiness/}
\BIBentrySTDinterwordspacing

\bibitem{10.1145/3563327}
\BIBentryALTinterwordspacing
R.~Bavishi, H.~Joshi, J.~Cambronero, A.~Fariha, S.~Gulwani, V.~Le, I.~Radi\v{c}ek, and A.~Tiwari, ``Neurosymbolic repair for low-code formula languages,'' \emph{Proc. ACM Program. Lang.}, vol.~6, no. OOPSLA2, Oct. 2022. [Online]. Available: \url{https://doi.org/10.1145/3563327}
\BIBentrySTDinterwordspacing

\bibitem{CodeCompose}
\BIBentryALTinterwordspacing
V.~Murali, C.~Maddila, I.~Ahmad, M.~Bolin, D.~Cheng, N.~Ghorbani, R.~Fernandez, N.~Nagappan, and P.~C. Rigby, ``Ai-assisted code authoring at scale: Fine-tuning, deploying, and mixed methods evaluation,'' \emph{Proc. ACM Softw. Eng.}, vol.~1, no. FSE, Jul. 2024. [Online]. Available: \url{https://doi.org/10.1145/3643774}
\BIBentrySTDinterwordspacing

\bibitem{Dunay2024FSE-industry}
\BIBentryALTinterwordspacing
O.~Dunay, D.~Cheng, A.~Tait, P.~Thakkar, P.~C. Rigby, A.~Chiu, I.~Ahmad, A.~Ganesan, C.~Maddila, V.~Murali, A.~Tayyebi, and N.~Nagappan, ``Multi-line ai-assisted code authoring,'' in \emph{Companion Proceedings of the 32nd ACM International Conference on the Foundations of Software Engineering}, ser. FSE 2024.\hskip 1em plus 0.5em minus 0.4em\relax New York, NY, USA: Association for Computing Machinery, 2024, p. 150–160. [Online]. Available: \url{https://doi.org/10.1145/3663529.3663836}
\BIBentrySTDinterwordspacing

\bibitem{SQLCompose}
C.~Maddila, N.~Ghorbani, K.~Jabre, V.~Murali, E.~Kim, P.~Thakkar, N.~P. Laptev, O.~Harman, D.~Hsu, R.~Abreu, and P.~C. Rigby, ``Ai-assisted sql authoring at industry scale,'' in \emph{Foundations of Software Engineering Industry Track}, 2025.

\bibitem{brown2024largelanguagemonkeysscaling}
\BIBentryALTinterwordspacing
B.~Brown, J.~Juravsky, R.~Ehrlich, R.~Clark, Q.~V. Le, C.~Ré, and A.~Mirhoseini, ``Large language monkeys: Scaling inference compute with repeated sampling,'' 2024. [Online]. Available: \url{https://arxiv.org/abs/2407.21787}
\BIBentrySTDinterwordspacing

\bibitem{Basili1999TSE}
V.~Basili, F.~Shull, and F.~Lanubile, ``Building knowledge through families of experiments,'' \emph{IEEE Transactions on Software Engineering}, vol.~25, no.~4, pp. 456--473, 1999.

\bibitem{jimenez2024swebenchlanguagemodelsresolve}
\BIBentryALTinterwordspacing
C.~E. Jimenez, J.~Yang, A.~Wettig, S.~Yao, K.~Pei, O.~Press, and K.~Narasimhan, ``Swe-bench: Can language models resolve real-world github issues?'' 2024. [Online]. Available: \url{https://arxiv.org/abs/2310.06770}
\BIBentrySTDinterwordspacing

\bibitem{swebenchverified}
\BIBentryALTinterwordspacing
(Accessed 2025) Introducing swe-bench verified. OpenAI. [Online]. Available: \url{https://openai.com/index/introducing-swe-bench-verified/}
\BIBentrySTDinterwordspacing

\bibitem{yang2025swesmithscalingdatasoftware}
\BIBentryALTinterwordspacing
J.~Yang, K.~Leret, C.~E. Jimenez, A.~Wettig, K.~Khandpur, Y.~Zhang, B.~Hui, O.~Press, L.~Schmidt, and D.~Yang, ``Swe-smith: Scaling data for software engineering agents,'' 2025. [Online]. Available: \url{https://arxiv.org/abs/2504.21798}
\BIBentrySTDinterwordspacing

\bibitem{miserendino2025swe}
S.~Miserendino, M.~Wang, T.~Patwardhan, and J.~Heidecke, ``Swe-lancer: Can frontier llms earn \$1 million from real-world freelance software engineering?'' \emph{arXiv preprint arXiv:2502.12115}, 2025.

\bibitem{ferrag2025llm}
M.~A. Ferrag, N.~Tihanyi, and M.~Debbah, ``From llm reasoning to autonomous ai agents: A comprehensive review,'' \emph{arXiv preprint arXiv:2504.19678}, 2025.

\bibitem{bouzenia2024repairagent}
I.~Bouzenia, P.~Devanbu, and M.~Pradel, ``{RepairAgent}: An autonomous, llm-based agent for program repair,'' \emph{arXiv preprint arXiv:2403.17134}, 2024.

\bibitem{zhang2024autocoderoverautonomousprogramimprovement}
\BIBentryALTinterwordspacing
Y.~Zhang, H.~Ruan, Z.~Fan, and A.~Roychoudhury, ``Autocoderover: Autonomous program improvement,'' 2024. [Online]. Available: \url{https://arxiv.org/abs/2404.05427}
\BIBentrySTDinterwordspacing

\bibitem{ruan2024specrover}
H.~Ruan, Y.~Zhang, and A.~Roychoudhury, ``{SpecRover}: {C}ode intent extraction via {LLM}s,'' \emph{arXiv preprint arXiv:2408.02232}, 2024.

\bibitem{chen2024coder}
D.~Chen, S.~Lin, M.~Zeng, D.~Zan, J.-G. Wang, A.~Cheshkov, J.~Sun, H.~Yu, G.~Dong, A.~Aliev \emph{et~al.}, ``Coder: Issue resolving with multi-agent and task graphs,'' \emph{arXiv preprint arXiv:2406.01304}, 2024.

\bibitem{liu2024marscode}
Y.~Liu, P.~Gao, X.~Wang, J.~Liu, Y.~Shi, Z.~Zhang, and C.~Peng, ``Marscode agent: Ai-native automated bug fixing,'' \emph{arXiv preprint arXiv:2409.00899}, 2024.

\bibitem{wang2024opendevin}
X.~Wang, B.~Li, Y.~Song, F.~F. Xu, X.~Tang, M.~Zhuge, J.~Pan, Y.~Song, B.~Li, J.~Singh \emph{et~al.}, ``Opendevin: An open platform for ai software developers as generalist agents,'' \emph{arXiv preprint arXiv:2407.16741}, 2024.

\bibitem{wang2025openhandsopenplatformai}
\BIBentryALTinterwordspacing
X.~Wang, B.~Li, Y.~Song, F.~F. Xu, X.~Tang, M.~Zhuge, J.~Pan, Y.~Song, B.~Li, J.~Singh, H.~H. Tran, F.~Li, R.~Ma, M.~Zheng, B.~Qian, Y.~Shao, N.~Muennighoff, Y.~Zhang, B.~Hui, J.~Lin, R.~Brennan, H.~Peng, H.~Ji, and G.~Neubig, ``Openhands: An open platform for ai software developers as generalist agents,'' 2025. [Online]. Available: \url{https://arxiv.org/abs/2407.16741}
\BIBentrySTDinterwordspacing

\bibitem{yang2024sweagentagentcomputerinterfacesenable}
\BIBentryALTinterwordspacing
J.~Yang, C.~E. Jimenez, A.~Wettig, K.~Lieret, S.~Yao, K.~Narasimhan, and O.~Press, ``Swe-agent: Agent-computer interfaces enable automated software engineering,'' 2024. [Online]. Available: \url{https://arxiv.org/abs/2405.15793}
\BIBentrySTDinterwordspacing

\bibitem{xia2023automated}
C.~S. Xia, Y.~Wei, and L.~Zhang, ``Automated program repair in the era of large pre-trained language models,'' in \emph{2023 IEEE/ACM 45th International Conference on Software Engineering (ICSE)}.\hskip 1em plus 0.5em minus 0.4em\relax IEEE, 2023, pp. 1482--1494.

\bibitem{xia2022less}
C.~S. Xia and L.~Zhang, ``Less training, more repairing please: revisiting automated program repair via zero-shot learning,'' in \emph{Proceedings of the 30th ACM Joint European Software Engineering Conference and Symposium on the Foundations of Software Engineering}, 2022, pp. 959--971.

\bibitem{barr2014plastic}
E.~T. Barr, Y.~Brun, P.~Devanbu, M.~Harman, and F.~Sarro, ``The plastic surgery hypothesis,'' in \emph{Proceedings of the 22nd ACM SIGSOFT International Symposium on Foundations of Software Engineering}, 2014, pp. 306--317.

\bibitem{xia2024automated}
C.~S. Xia and L.~Zhang, ``Automated program repair via conversation: Fixing 162 out of 337 bugs for \$0.42 each using {ChatGPT},'' in \emph{Proceedings of the 33rd ACM SIGSOFT International Symposium on Software Testing and Analysis}, 2024, pp. 819--831.

\bibitem{xia2024agentless}
C.~S. Xia, Y.~Deng, S.~Dunn, and L.~Zhang, ``Agentless: Demystifying llm-based software engineering agents,'' \emph{arXiv preprint arXiv:2407.01489}, 2024.

\end{thebibliography}

\end{document}